# AN ONTOLOGY FOR MODELLING AND SUPPORTING THE PROCESS OF AUTHORING TECHNICAL ASSESSMENTS

Khalil Riad Bouzidi, Phd student, khalil-riad.bouzidi@cstb.fr
Bruno Fies, Senior engineer, bruno.fies@cstb.fr
Marc Bourdeau, Deputy head of division, marc.bourdeau@cstb.fr
*Centre Scientifique et Technique du Bâtiment (CSTB), Sophia Antipolis, France*
Catherine Faron-Zucker, Senior lecturer, catherine.faron-zucker@unice.fr
Nhan Le Thanh, Professor, nhan.le-thanh@unice.fr
*Laboratoire I3S, Université de Nice Sophia Antipolis, CNRS, France*

**ABSTRACT**

In this paper, we present a semantic web approach for modelling the process of creating new technical and regulatory documents related to the Building sector. This industry, among other industries, is currently experiencing a phenomenal growth in its technical and regulatory texts. Therefore, it is urgent and crucial to improve the process of creating regulations by automating it as much as possible. We focus on the creation of particular technical documents issued by the French Scientific and Technical Centre for Building (CSTB), called Technical Assessments, and we propose services based on Semantic Web models and techniques for modelling the process of their creation.

**Keywords:** Ontology, Semantic Web, Knowledge Management, Building Industry, e-regulations.

## 1 INTRODUCTION

In this article, we address the general problem of the creation of regulatory documents and present an engineering overview to help staff in the authoring of new regulations. Our approach takes into account the constraints expressed in the existing statutory corpus. We chose to apply our work in the context of the French Scientific and Technical Centre for Building (CSTB) to help it in the writing of Technical Assessments. The issue is to specify how such assessment document is created and how to standardize its structure using models and adaptive semantic services.
A Technical Assessment (in French: Avis Technique ou ATec) is a document containing technical information on the usability of a product, material, component or element of construction, which has an innovative character. We chose this Technical Assessment as a study model because CSTB has the mastership and a wide experience in these kinds of technical documents. We were able to lead interviews on the CSTB site of Sophia-Antipolis with experts directly involved in the drafting of technical assessments.
There are multiple users of Technical Assessments:

- Firms working on tender responses and looking for products validated by an Atec;
- Manufacturers working on the technical documentation of a new product in order to apply for an ATec;
- Insurers who need to verify that a product is implemented according to its domain of use;
- Craftsmen looking for a product that correspond to a subfamily and a particular job domain.

The study shows that the technical assessment consists of three parts: (I) an overview and identification of the assessment, (II) the technical assessment itself, formulated by a group of experts from CSTB, and (III) the technical document of the product or process which must be delivered in the technical assessment. We are particularly interested in assisting the creation of technical documents, using entry forms that their enchainment will consider at a time the information modelled in ontology and the entries to be made by the user. In the next section, we describe the process of editing technical assessment; in section 3, we detail our approach to support the creation of technical documents. In section 4, we introduce a tool developed to validate our approach. Finally, we conclude in section 5.

## 2 PROCESS OF EDITING TECHNICAL ASSESSMENTS

A Technical Assessment is drafted at the request of an industrial. The industrial starts by sending a request for technical assessment to relevant departments within the CSTB. Then the CSTB instructors transmit to the industrial the resolution of technical assessment and a form for developing the technical document. It is a Word file containing chapters, text and instructions. This form contains all information necessary for the process; the products or materials are studied by a specialized group who will be responsible for delivering the technical assessment. The industrial is supposed to fill in the form and send a complete document to CSTB, which in turn transmits an estimate of the technical assessment to the industrial.

Once the estimate is accepted by the industrial, the CSTB instructors will check the contents of the technical document and can ask the industrial to justify some elements shown in his case by experimental tests that will serve as proof of the resistance of the product to various external risks. Site and / or factory visits can be planned to ensure smooth functioning of the manufacturing processes and / or product installation. To be valid the technical document must eventually include a detailed description of the field and conditions of use to the applicant, the requirements for implementation, the experimental results, and the references of use.

Once the file is complete, CSTB instructors transfer it to the relevant specialist group. The assignment of this group is to verify the compliance of information according to current standards, and to issue an assessment. Then CSTB instructors publish the assessment on CSTB website and transfer a copy to the industrial.

The process above described is ideal. In practice, it is not fully followed. Through our discussions with CSTB experts, it appears clearly that the industrial fails to properly inform the weft of the technical document. A dialogue between CSTB and the applicant too many exchanges are necessary before reaching an acceptable version of the technical file. Without CSTB's accompaniment, most ATec applicants fail to properly complete their technical documents. On its side, CSTB is not able to afford to mobilize resources outside some estimate. The industrial's process is poorly anticipated and considered too administrative. As a direct consequence, this leads to a long casework, about 6 to 8 months. The role of CSTB is limited to verify the contents of the document and not to assist the industrial. The latter must take on the task to fill in the weft of the technical dossier correctly before returning it to CSTB.

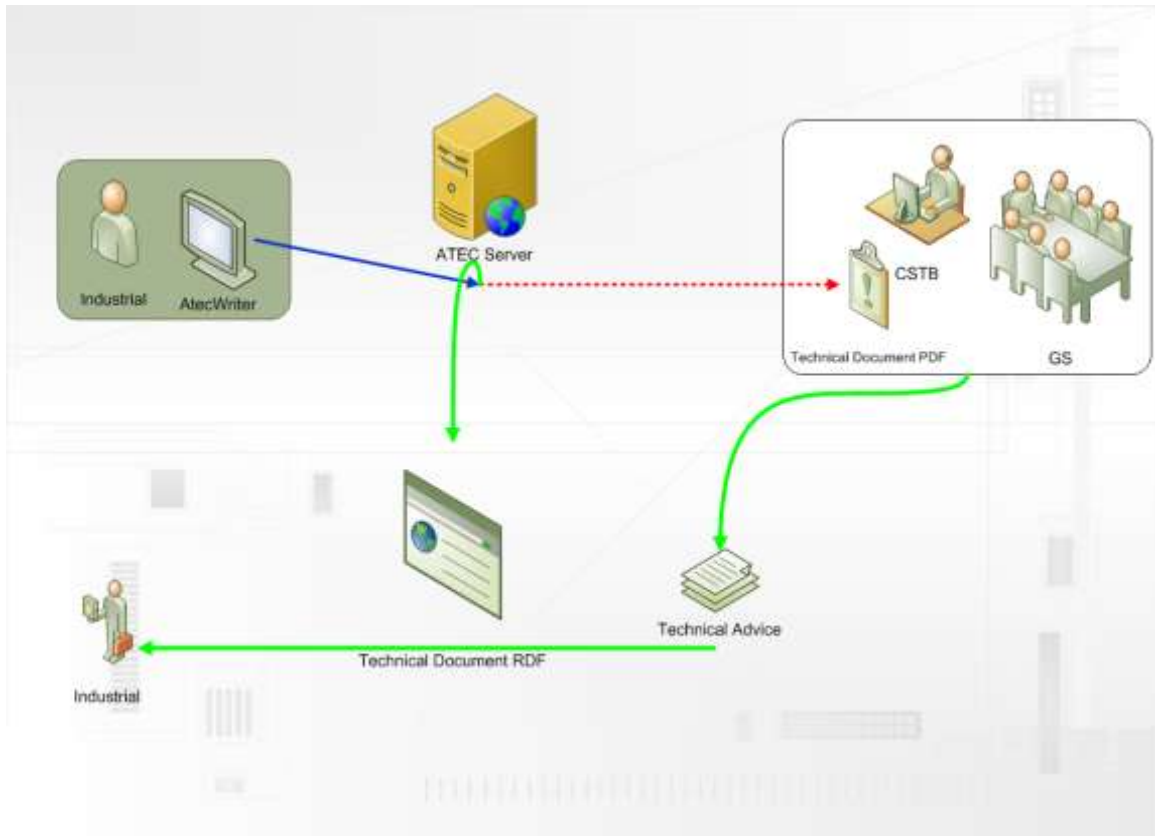

Figure 1: The process of creating technical assessment

The slowness of the process is accentuated by the absence of a common format for representing various sources of data that CSTB instructors will be led to examine in the technical document. The technical document is a long Word form sent by CSTB instructors to the industrial to certify their products / processes / materials. This file to be completed is generic for a particular domain. We have realized in our study of ATec in the photovoltaic area of that the technical document to be completed contains information common to all photovoltaic modules, that is to say, information relative to all the products that can be used in the field of photovoltaic. Industrial thus have each to fill the only parts concerning them. But industrials are struggling to understand this file and inquire properly. As a result, the CSTB's instructors receive a badly informed technical document.

Following this analysis, we concluded that the process of writing technical documents would benefit from being modelled and supported by tools to guide this creation by identifying at the same time the parts to be informed and the expected information.

## 3  MODELLING THE PROCESS OF CREATING TECHNICAL AND REGULATORY DOCUMENTS

In continuation of the work performed at CSTB (Yurchyshyna 2009), on building ontology and checking conformity of construction projects (Gehre and Katranuschkov 2005), we propose to model the process of creating the technical dossier using an ontology-based approach to build a semantically rich model of the weft of the technical document.

### 3.1  Construction of the technical document ontology

We have studied the technical documents issued by the PV group within CSTB with the objective to to build an ontology of the domain based on the terms identified from the weft of the technical document. Many solutions have been proposed for the manual construction of ontologies, notably (Gruninger and Fox 1995), (Ushold and King 1995), Methontology (Gomez-Perez et al. 2003), and On-To-Knowledge (Staab et al. 2001). The common idea in these solutions is to reuse as much as possible existing ontologies related to the domain to model. This is the approach that we chose to adopt: we do not address the problems of NLP[1] , we only build an ontology from experts' interviews and by reusing existing ontological resources.

Specifically, in our work, with the help of PV experts, we began to identify what knowledge to represent and with which degree of accuracy. From the weft of the technical document, we extracted the terms corresponding to different elements or components of the product to certify . Then, we associated concepts to those terms. Our first result is an ontology where each concept considered relevant to the creation of technical documents has been identified and modelled. We have compared our ontology with the terms of the building thesaurus developed by CSTB for its REEF product. The idea of reusing the CSTB thesaurus is to have a controlled vocabulary and to potentially link / integrate the ATec in the REEF (Bus et al. 2009).

By doing so, we adopted an approach similar to that of (Hernandez et al. 2006): reusing the domain thesaurus that required heavy design efforts for the development of new resources with a higher formal level. This principle is interesting insofar as it avoids building a new ontology from scratch. The design of ontologies from thesauri has the advantage of considering all the terms identified by experts as being representative of the domain.

We adapted this approach to our case study by reusing part of the transformation process proposed by (Hernandez et al. 2006) and merging the resulting ontology with our ontology of technical document. However, the resulting ontology is light, with a minimum of semantics; it contains general concepts of the building industry and it lacks of specific terms of the Photovoltaic industry. To respond to our problem we need a specific concept issue from technical documents.

Our approach reflects this in three steps: (I) Extracting hierarchical relationships explicit in the REEF thesaurus, (II) Removing redundancy in hierarchical relations of the thesaurus, (III) Merging REEF ontology resulting of steps (I) and (II) with our ontology of technical documents.

### 3.1.1 Extracting concepts and relations

For this task, we used the REEF resources, containing a majority of entities in the building field. For example, the core concept of "étanchéité" (sealing) is connected to several more specific concepts (narrower terms) as « joint d'étanchéité » (watertight joint) or « étanchéité à l'air » (airtightness). It is also tied by a relationship "Broad" (broader term) to the concept of « calfeutrage » (draughtproofing).

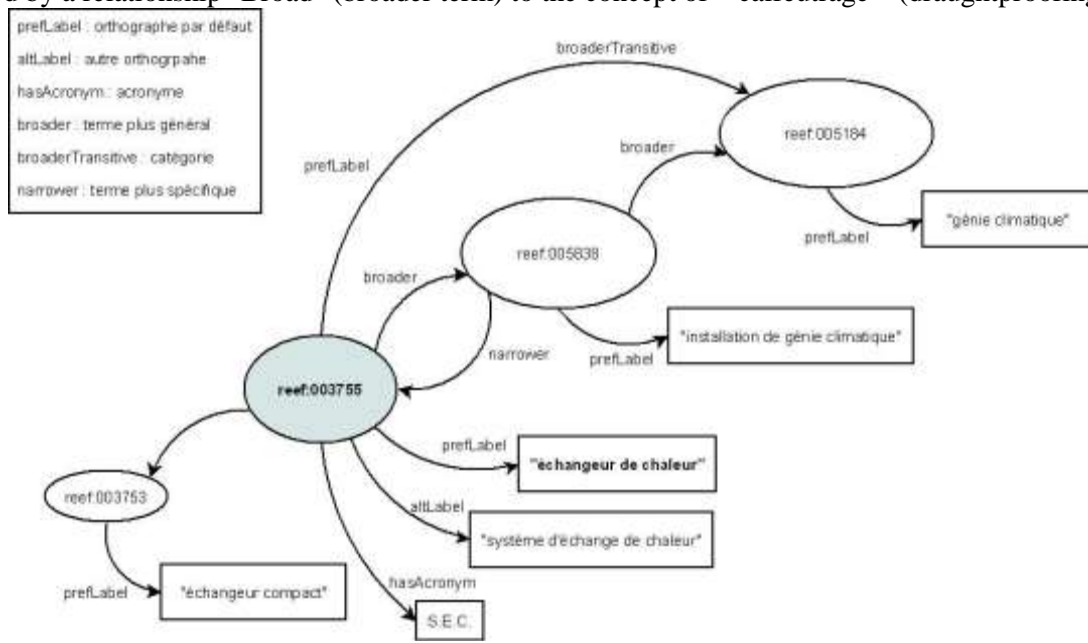

Figure 2: Extract of REEF thesaurus

---

[1] Natural language processing

First, we organized the concepts of the REEF thesaurus hierarchically based on the relation "subclassOf" in the conceptual schema of the ontology. Some hierarchical relationships between concepts are directly derived from explicit links in the thesaurus. To extract the relationships of the thesaurus, the relations "narrower" and "broader" are taken into account. These are selected as potential candidates for "subclassOf" relationships between a concept and the concept to which the related term in the thesaurus.

If not formalized, redundancies in the structure of the thesaurus may exist. Our ontology can be confronted with transitive relations of the inference type: if A is a subclassOf B and B is a subclassOf C, then A is a subclassOf C. A, B, C are concepts the hierarchical relationship between A and C need not be explicit. To remove this type of redundant relationships, we checked the relevance of each relation "is a subclass of". Hernandez offers an analysis through graph theory.

### 3.1.2 Merging REEF ontology and ontology of technical document

Once we have established the REEF ontology we began the part of integration consisting in bringing closer the REEF ontology concepts with the concepts of the ontology of the Technical document (OntoDT) that have the same label. We aim to clarify the semantics that is vague (soft) around the REEF terms describing the technical document by restricting it to the regulatory aspect of the PV field. To achieve this integration we chose to merge the REEF ontology and those of technical document (OntoDT). The resulting ontology unifies and replaces the original ontology

The most common approaches for merging ontologies use union or intersection to connect the resulting ontology to the original ontologies. In the union approach, the resulting ontology contains the union of entities coming from the original ontologies and suppose resolved the differences of representation of the same concept. In the intersection approach, the resulting ontology contains only the common parts of the original ontologies. We adopted the intersection approach: once the merge is completed, the resulting ontology is the intersection between the REEF ontology and the ontology of technical document

Several approaches and systems for merging ontologies have been proposed, including PROMPT (Noy et al. 2000), Chimaera (Mcguinness et al. 2000), OntoMerge (Dou et al. 2002). We chose PROMPT as a tool for merging the REEF ontology and the technical document ontology as it creates a complete ontology. It identifies a set of operations for ontology merging (fusion classes, merging of links) and a set of possible conflicts resulting from the application of these operations (name conflicts, redundancy in the class hierarchy).

As a result, our ontology of technical document has 50 classes and 26 properties formalised in the OWL[2] Lite language. 35% of these classes are created from REEF terms. The remaining 65% are concepts more specific than those of the REEF thesaurus, which contains general concepts of the building industry. In its current state, it lacks specific terms relative to a particular field (Photovoltaic). However, it remains in constant evolution

### 3.2 Modelling the technical document

We use our ontology of the technical document to represent the technical document. Our approach is to model the semantics conveyed by the technical document to a formal interpretable knowledge. We changed the formatting of the technical document as a set of forms interconnected with each other, each one relative to a component of the product to be described. Each form contains components that represent the nomenclature of the same product, the component itself may be formed of one or more elements.

For example, a PV glass polymer module (a component) enters into the composition of a PV panel (the product). This glass polymer module is composed of several elements: polymer film, photovoltaic cell, etc. (elements).

To model this composition, each concept representing a component or an element or a product is defined by an axiom. Axioms are used in the definition of an ontological class, they are of the form A ⊑ B, where A is a primitive concept (Product) and B description composed concept. For instance, Figure 3 provides the definition of class VerrePolymere in the OWL language: it is a subclass of class

---
[2] http://www.w3.org/TR/owl-ref/

ModulePhotoVoltaic and of a class defined as the intersection of the classes of the instances having components of class CableElectrique, Cadre, Cellule PhotoVoltaique, etc.

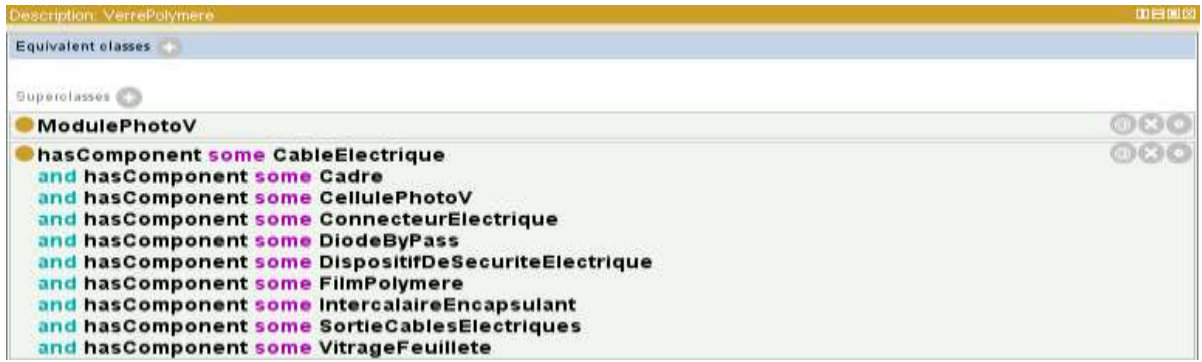

```xml
<owl:Class rdf:about="#VerrePolymere">
  <rdfs:subClassOf rdf:resource="#ModulePhotoV"/>
  <rdfs:subClassOf>
    <owl:Class>
      <owl:intersectionOf rdf:parseType="Collection">
        <owl:Restriction>
          <owl:onProperty rdf:resource="#hasComponent"/>
           <!-- CableElectrique -->
          <owl:someValuesFrom rdf:resource="http://www.cstb.fr/reef/#01573"/>
        </owl:Restriction>
        <owl:Restriction>
          <owl:onProperty rdf:resource="#hasComponent"/>
          <!-- Cadre -->
          <owl:someValuesFrom rdf:resource=" http://www.cstb.fr/reef/#01593"/>
        </owl:Restriction>
              ...
      </owl:intersectionOf>
    </owl:Class>
  </rdfs:subClassOf>
</owl:Class>
```

Figure 3: Example of a defined concept

We have described the interdependencies between the different concepts modelling the interactions or interdependencies between parts of PV modules. These have a number of intrinsic characteristics (length, weight, manufacturer, etc.) represented in our modelling by properties attached to the corresponding concepts.
Once the industrial fills in the form relative to a concept with all the elements relative to its definition, they are stored in an RDF[3] annotation file. By doing so, we acquire an interoperable representation of a technical document, reusable in other systems. The use of the RDF model will allow us later to check the conformity of the technical document with the standards of the photovoltaic domain.

### 3.3 Modelling the process of writing a technical document

The question which now arises, is how to model the process itself of filling the technical document, in order to produce a dynamic sequence of forms to fill. The dynamicity stands in the fact that the forms to fill in depend on the way previous forms are filled in: we want to adjust to the information provided without requiring him to fill irrelevant parts of the technical document.
Our approach is to ask first of high-level information (name the type of product, etc.) then browse the explicit dependency rules in the ontology to seek all information required for a product. Each information provided by the industrial will be confronted to the ontology and will determine the next interaction. The industrial has to choose in a first form a product among all those concerned with the photovoltaic field. From that first choice, based on our ontology, we determine the list of components used in its manufacture and we offer a list of corresponding forms.

---

[3] http://www.w3.org/RDF/

```
SELECT ?Composant WHERE{
 ?x rdfs:subClassOf ?y
 FILTER( ?x=dt:ComposantName)
 ?y owl:intersectionOf ?z
 ?z rdf:rest*/rdf:first ?f
 ?f owl:onProperty ?p
 ?f owl:someValuesFrom ?Composant}
```

Figure 4: Extract of the SPARQL query pattern

The dynamicity of the sequence of forms relies upon a SPARQL[4] query pattern presented in Figure 4 that we instanciate to query the ontology to determine each next form. More precisely, by using this query template, we query a product on its composition by questioning its definition. The concepts involved in its definition are returned by the query and are so much information that the industrial must provide through entry forms generated on the fly. The chaining of forms thus depends on the query results: each form corresponds to one or more elements of the result.

For example, the query below searches the ontology on the definition of the concept "VerrePolymere".

```
SELECT ?Composant WHERE{
 ?x rdfs:subClassOf ?y
 FILTER( ?x=dt:VerrePolymere)
 ?y owl:intersectionOf ?z
 ?z rdf:rest*/rdf:first ?f
 ?f owl:onProperty ?p
 ?f owl:someValuesFrom ?Composant}
```

The result is that "VerrePolymere" is a module which has as components a "Cadre", a "CellulePhotoV", etc.
`Result : [Cadre, CellulePhotoV, FilmPolymere, VerreInterieur…]`

At each step of the process of writing a technical document, we display to the industrial a form for entering information related to a concept belonging to the result of such a query. If this concept is itself defined, the same query template is instanciated with a new query in order to provide the Industrial with new forms matching with the components involved in the definition of the current concept. The same query pattern is recursively instanciated until reaching terminal concepts, i.e. primitive concepts (with no definition).

As a result, the industrial browses thereby transparently into our ontology to complete all components of its product by filling out the forms provided; only relevant questionnaires are displayed.

## 4   SUPPORT TOOL FOR THE CREATION OF TECHNICAL DOCUMENTS

To validate our model, we developed a tool to assist in the creation of technical documents. This application has been developed in J2EE. It conforms to the three-tier architecture.

The first third consists of a web page on the client side (GUI) that supports various forms requiring to be filled in by the industrial.

The second third on server side is a servlet containing business code that interacts with the GUI part and the business part. It is connected to the CORESE[5] semantic engine (Corby et al. 2004) to query the ontological knowledge that represents of the technical documents and generates the RDF model of the filled technical document.

The last third is relative to the data. It includes the ontology of the technical document and information of the product stored as RDF annotations.

Our tool provides the industrials with a rich interface, interactive and easy of handling. At the launch of the application, a first general form allows the industrial to define the context of its technical document. It is at this first level of choice in a dropdown lists. The choices correspond to concepts of the ontology of technical documents whose labels correspond to the vocabulary defined in the thesaurus of the REEF. Depending on the choices made by the industrial in this first level, the system adapts and goes through the ontology to seek to progressively more specific information. These sequences of forms are orchestrated through SPARQL queries that run on the ontology of the technical document.

---

[4] http://www.w3.org/TR/rdf-sparql-query/
[5] http://www-sop.inria.fr/edelweiss/software/corese/

At the end of the sequence of forms, two files are generated: a readable file in format PDF / HTML containing the information filled in by the industrial, and a semantic description of the document in RDF. This will be exploited to help in writing the technical assessments itself, by automatically checking the conformity of the information in the technical document with the regulatory texts of the domain.

We developed a first version of this tool and we have submitted to functional tests with photovoltaic instructors of the CSTB. This allowed us of validate the pertinence of our approach.

## 5  CONCLUSION

In this paper, we have presented an innovative approach to support the authoring of technical documents by industrials, and thus get technical assessment on their products more quickly. Our approach is based on the construction of an ontology relative to the concerned domain which concepts represent different elements of the technical document This knowledge is exploited through a semantic query pattern to establish a dynamic chaining of forms. The instanciation of this query pattern provides a dynamic chaining of forms, coherent with user entries, where no component of the product is missing, where the entire elements specific to the product are shown.

We have developed a tool to assist the creation of technical documents based on such a dynamic chaining of forms. It is based on the languages and techniques of the Semantic Web, ensures the interoperability of knowledge, and manipulated the results

A first version exists which allowed us to validate our approach. We now need to enrich the ontology to compare our tool to a real use case.

## REFERENCES

Bus, N. et al. (2009). "Reef sémantique, Diffusion et application des textes technico-réglementaires, Accompagnement des pouvoirs publics dans la rédaction des textes officiels, Livrable 4, CSTB.

Corby, O. et al. (2004). "Querying the SemanticWeb with Corese Search Engine" *Proc. of the 16th Eureopean Conference on Artificial Intel- ligence, ECAI, IOS Press,* 705-709.

DOU, D. et al. (2002) "Ontology translation by ontology merging and automated reasoning" *Proceedings of EKAW Workshop on Ontologies for Multi-Agent Systems*

Gehre, A. et al. (2005). "Towards semantic interoperability in virtual organisations" *in Proc. of the 22nd Conference on Information Technology in Construction, Dresden, Germany*.

Gómez-Pérez, A. et al. (2004) " Ontological Engineering: With Examples from the Areas of Knowledge Management, E-commerce and the Semantic Web" Springer Verlag *Advanced Information and Knowledge Processing*

Grüninger, M. et al. (1995) "Methodology for the Design and Evaluation of Ontologies" *Proceedings of the IJCAI-95 Worshop on Basic Ontological Issues in Knowledge Sharing*

Guarino, N. (1997). "Understanding building, and using ontologies". *International Journal of Human and Computer Studie*.

Hernandez, N. et al (2006) "TtoO: une méthodologie de construction d'ontologie de domaine à partir d'un thésaurus et d'un corpus de référence" *Rapport interne IRIT/RR*

Mcguinness, D. et al. (2000) "The chimaera ontology environment" *Proceedings of the Seventeenth National Conference on Artificial Intelligence and Twelfth Conference on on Innovative Applications of Artificial Intelligence,* Austin, Texas, USA, 1123–1124.

Noy, N. et al. (2000) "Prompt: Algorithm and tool for automated ontology merging and alignment" *Proceedings of the Seventeenth National Conference on Artificial Intelligence and Twelfth Conference on on Innovative Applications of Artificial Intelligence,* Austin, Texas, USA, 450-455

Staab, S. et al. (2001) "Knowledge Processes and Ontologies." *IEEE Intelligent Systems 16 (1)*, 26-34

Uschold, M. et al. (1995) "Towards a Methodology for Building Ontologies" *Proceedings of the IJCAI-95 Worshop on Basic Ontological Issues in Knowledge Sharing*.

Yurchyshyna, A. (2008). "Towards the Knowledge Capitalisation and Organisation in the Model of Conformity-Checking Process in Construction" *In Proc. of the 12th International Conference on Knowledge-Based and Intelligent Information & Engineering Systems (KES 2008), Zagreb, Croatia.*
Yurchyshyna, A. (2009). "Modélisation du contrôle de conformité en construction : une approche ontologique" *Thèse de l'université de Nice Sophia Antipolis, France*